\documentclass[prb,twocolumn,showpacs,aps,superscriptaddress,floatfix]{revtex4}
\usepackage{amsmath}
\usepackage{amssymb}
\usepackage{amsfonts}
\usepackage{bm}
\usepackage{graphicx}
\usepackage{color}
\usepackage[svgnames]{xcolor}
\usepackage{ulem}
\usepackage{hyperref}
\usepackage{verbatim}

\begin{document}

\title{Many-body effects in twisted bilayer graphene at low twist angles}

\author{A.O. Sboychakov}
\affiliation{Theoretical Quantum Physics Laboratory, RIKEN, Wako-shi,
Saitama, 351-0198, Japan}
\affiliation{Institute for Theoretical and Applied Electrodynamics, Russian
Academy of Sciences, Moscow, 125412 Russia}

\author{A.V. Rozhkov}
\affiliation{Theoretical Quantum Physics Laboratory, RIKEN, Wako-shi,
Saitama, 351-0198, Japan}
\affiliation{Institute for Theoretical and Applied Electrodynamics, Russian
Academy of Sciences, Moscow, 125412 Russia}
\affiliation{Moscow Institute for Physics and Technology (State
University), Dolgoprudnyi, 141700 Russia}
\affiliation{Skolkovo Institute of Science and Technology, Skolkovo
Innovation Center 3, Moscow 143026, Russia}

\author{A.L. Rakhmanov}
\affiliation{Theoretical Quantum Physics Laboratory, RIKEN, Wako-shi,
Saitama, 351-0198, Japan}
\affiliation{Institute for Theoretical and Applied Electrodynamics, Russian
Academy of Sciences, Moscow, 125412 Russia}
\affiliation{Moscow Institute for Physics and Technology (State
University), Dolgoprudnyi, 141700 Russia}
\affiliation{Dukhov Research Institute of Automatics, Moscow, 127055
Russia}

\author{Franco Nori}
\affiliation{Theoretical Quantum Physics Laboratory, RIKEN, Wako-shi,
Saitama, 351-0198, Japan}
\affiliation{Department of Physics, University of Michigan, Ann Arbor, MI
48109-1040, USA}

\begin{abstract}
We study the zero-temperature many-body properties of twisted bilayer
graphene with a twist angle equal to the so-called `first magic angle'. The
system low-energy single-electron spectrum consists of four (eight, if spin
label is accounted) weakly-dispersing partially degenerate bands, each band
accommodating one electron per Moir{\'{e}} cell per spin projection. This
weak dispersion makes electrons particularly susceptible to the effects of
interactions. Introducing several excitonic order parameters with
spin-density-wave-like structure, we demonstrate that (i)~the band
degeneracy is partially lifted by the interaction, and (ii)~the details of
the low-energy spectrum becomes doping-dependent. For example, at or near
the undoped state, interactions separate the eight bands into two quartets
(one quartet is almost filled, the other is almost empty), while for two
electrons per Moir\'{e} cell, the quartets are pulled apart, and doublets
emerge. When the doping is equal to one or three electrons per cell, the
doublets split into singlets. Hole doping produces similar effects. As a
result, electronic properties (e.g., the density of states at the Fermi
energy) demonstrate oscillating dependence on the doping concentration.
This allows us to reproduce qualitatively the behavior of the conductance
observed recently in experiments
[Cao et al., Nature {\bf 556}, 80 (2018)].
Near half-filling, the electronic spectrum loses hexagonal symmetry
indicating the appearance of a many-body nematic state.
\end{abstract}


%
%
%
%
%
%
%
%
%
%

\date{\today}

\maketitle

\section{Introduction}
The search for broken-symmetry phases in graphene bilayer systems remains
an active research
area~\cite{ourBLGreview2016}.
Theorists have studied a variety of possibilities, such as
antiferromagnetism~\cite{haritonov_afm2012,AkzyanovAABLG2014,Honerkamp,
lemonic_rg_nemat_long2012,Nilsson2006,PrbROur,PrbOur,lang_af_hubb2012,
PrlOur},
superconductivity~\cite{gonzalez_rotenberg_supercond_mono2010,
kagan_supercond_blg2015,kagan_supercond_jetp2014,
gonzalez_supercond_mono2001,Mohammad},
excitons~\cite{LozovikSokolik,LozovikPLA2009,AkzyanovAABLG2014,
our_prl_twist_bias2018},
as well as more exotic
states~\cite{lemonic_rg_nemat_long2012,zhang_quant_hall2012}.
Unfortunately, experimentally, the broken symmetry phases are rare
celebrities in graphene systems, except, perhaps, AB~bilayer graphene, for
which numerous
experiments~\cite{freitag2013,Maher2013,Bao2012,Elferen2012,Freitag2012,
Velasco2012}
provide evidence of low-temperature non-superconducting order. It appears,
however, that the situation in this field has changed: in recent
experiments~\cite{twist_exp_insul2018,twist_exp_sc2018}
both superconductivity and many-body insulating states were detected in
doped samples of twisted bilayer graphene (TBLG) whose twist angles
$\theta$ are close to the so-called `first magic angle'
$\theta_c \approx 1.1^\circ$.
The dependence of the conductance $\sigma$, as a function of doping $n$,
showed several pronounced minima: at
$n=0$
(undoped state), at
$n=\pm n_s/2$,
and at
$n=\pm n_s$
(the doping level
$n=n_s$
corresponds to one electron per spin projection per layer per supercell,
or, equivalently, four electrons per supercell). In some samples,
additional minima were observed
at~\cite{extended_cao}
$n=\pm 3 n_s/4$,
and
at~\cite{weak_minimum}
$n= n_s/4$.
The purpose of this paper is to offer a theoretical explanation to these
remarkable findings.

Our reasoning relies on the peculiar band structure of TBLG at small twist
angles: for
$\theta \leq \theta_c$,
the low-energy single-electron spectrum is dominated by four (eight, if
spin degeneracy is accounted) bands with almost no
dispersion~\cite{extended_cao_dc},
and the Fermi surface is present even at zero
doping~\cite{ourTBLG}
(provided that the interaction effects are neglected). The single-electron
density of states (DOS) of these bands offers a simple
explanation~\cite{twist_exp_insul2018}
for the conductance minima at
$n = \pm n_s$.
As for the minima in the interval
$-n_s < n < n_s$,
such single-body reasoning fails to explain them, and a many-body formalism
is necessary. Indeed, the flatness and degeneracy of the low-energy bands
make them particularly susceptible to the interaction effects. To account
for the latter, we use a mean-field approach. A simple single-site
spin-density wave (SDW) order parameter is sufficient to reproduce the
minimum at
$n=0$:
in energy space, such an order parameter splits the eight bands into two
quartets, one quartet is almost filled, the other is almost empty, with
drastically reduced DOS at the Fermi level. To explain the behavior of
$\sigma (n)$
at other $n$'s, the quartets must be split further (into doublets and
singlets), which requires more complex SDW order parameters. The resultant
formalism captures qualitatively the dependence of $\sigma$ versus doping
reported in
Ref.~\onlinecite{twist_exp_insul2018}.
In addition, our calculations demonstrate that for sufficiently large
doping the so-called electronic nematicity may be stabilized.

The paper is organized as follows. The basic facts about the TBLG geometry
are outlined in
Sec.~\ref{sec::geometry}.
The studied model is formulated in
Sec.~\ref{sec::model}.
The mean field approximation is applied to the model in
Sec.~\ref{sec::MF_calcs}.
Section~\ref{sec::discussion}
is dedicated to the discussions of the presented results,
while the conclusions are formulated in
Sec.~\ref{sec::conclusion}.

\section{Geometry of twisted bilayer graphene}
\label{sec::geometry}

To introduce the notation, let us start with a brief review of basic TBLG
geometrical facts. More details can be found in
Refs.~\onlinecite{dSPRB,MeleReview,ourBLGreview2016}.
A graphene monolayer has a hexagonal crystal structure consisting of two
triangular sublattices
${\cal A}$
and
${\cal B}$.
The coordinates of atoms in layer $1$ on sublattice
${\cal A}$
are
\begin{eqnarray}
\mathbf{r}_{\mathbf{n}}^{1{\cal A} }
=
\mathbf{r}_{\mathbf{n}}^{1}\equiv n\mathbf{a}_1+m\mathbf{a}_2,
\end{eqnarray}
where
$\mathbf{n}=(n,\,m)$
is an integer-valued vector,
\begin{eqnarray}
\mathbf{a}_{1,2}=a(\sqrt{3},\mp1)/2
\end{eqnarray}
are the primitive vectors,
$a=2.46$\,\AA\
is the lattice constant of graphene. The coordinates of atoms on sublattice
${\cal B}$
are
\begin{eqnarray}
\mathbf{r}_{\mathbf{n}}^{1{\cal B} }
=
\mathbf{r}_{\mathbf{n}}^{1}+\bm{\delta},
\end{eqnarray}
where
\begin{eqnarray}
\bm{\delta}=a(1/\sqrt{3},0).
\end{eqnarray}
Atoms in layer $2$ are located at
\begin{eqnarray}
\mathbf{r}_{\mathbf{n}}^{2{\cal B} }
=
\mathbf{r}_{\mathbf{n}}^{2}
\equiv
d\mathbf{e}_z+n\mathbf{a}_1'+m\mathbf{a}_2',
\quad
\mathbf{r}_{\mathbf{n}}^{2{\cal A} }
=
\mathbf{r}_{\mathbf{n}}^{2}-\bm{\delta}',
\end{eqnarray}
where
$\mathbf{a}_{1,2}'$
and
$\bm{\delta}'$
are the vectors
$\mathbf{a}_{1,2}$
and
$\bm{\delta}$,
rotated by an angle $\theta$. The unit vector along the $z$-axis is
$\mathbf{e}_z$,
the inter-layer distance is
$d=3.35$\,\AA.
The limiting case
$\theta=0$
corresponds to the AB stacking.

If the twist angle satisfies
\begin{eqnarray}
\cos\theta
=
\frac{3m_0^2+3m_0r+r^2/2}{3m_0^2+3m_0r+r^2},
\end{eqnarray}
where
$m_0$
and $r$ are co-prime positive integers, a superstructure emerges, and a
TBLG sample splits into a periodic lattice of finite supercells. The number
of graphene unit cells inside a supercell is
\begin{eqnarray}
N_{\rm sc}=(3m_0^2+3m_0r+r^2)/g
\end{eqnarray}
per layer, where
$g=1$
if
$r\neq3n$,
or
$g=3$
otherwise. 

The reciprocal lattice primitive vectors for the layer~1 are denoted by
$\mathbf{b}_{1,2}$,
for layer~2 they are
$\mathbf{b}_{1,2}'$.
In layer~1 we have
\begin{eqnarray}
\mathbf{b}_{1,2}=(2\pi/\sqrt{3},\mp 2\pi )/a,
\end{eqnarray}
while
$\mathbf{b}_{1,2}'$
are connected to
$\mathbf{b}_{1,2}$
by rotating on angle $\theta$. 

When the superlattice is present, the primitive reciprocal vectors for the
superlattice can be defined. We denote them as
$\bm{{\cal G}}_{1,2}$.
For these vectors, the following identities in the reciprocal space are
valid:
\begin{eqnarray}
\mathbf{b}_1'=\mathbf{b}_1+r(\bm{{\cal G}}_{1}+\bm{{\cal G}}_{2}),
\quad
\mathbf{b}_2'=\mathbf{b}_2-r\bm{{\cal G}}_{1},
\end{eqnarray}
if
$r\neq3n$,
or
\begin{eqnarray}
\mathbf{b}_1'=\mathbf{b}_1+r(\bm{{\cal G}}_{1}+2\bm{{\cal G}}_{2})/3,
\\
\mathbf{b}_2'=\mathbf{b}_2-r(2\bm{{\cal G}}_{1}+\bm{{\cal G}}_{2})/3,
\end{eqnarray}
otherwise. The Brillouin zone of the superlattice is hexagonal-shaped. It
can be obtained by
$N_{\rm sc}$-times
folding of the Brillouin zone of the layer $1$ or $2$. Two non-equivalent
corners of the reduced Brillouin zone,
$\mathbf{K}_1$
and
$\mathbf{K}_2$,
can be expressed via vectors
$\bm{{\cal G}}_{1,2}$
as
\begin{eqnarray}
\mathbf{K}_1=(\bm{{\cal G}}_{1}+2\bm{{\cal G}}_{2})/3,
\quad
\mathbf{K}_2=(2\bm{{\cal G}}_{1}+\bm{{\cal G}}_{2})/3.
\end{eqnarray}

\section{Model Hamiltonian}
\label{sec::model}

\subsection{Single-electron term}

We investigate the tight-binding model for
$p_z$
electrons in the TBLG at small doping $n$. The Hamiltonian is
\begin{eqnarray}
{\hat H} = {\hat H}_0 + {\hat H}_{\rm int},
\end{eqnarray}
where
${\hat H}_{\rm int}$
is the electron-electron interaction, and a single-electron term equals to
\begin{eqnarray}
\label{eq::single_H}
{\hat H}_0\!=\!\sum_{{i\mathbf{n}j\mathbf{m}\atop ss'\sigma}}\!
	t(\mathbf{r}_{\mathbf{n}}^{is};\mathbf{r}_{\mathbf{m}}^{js'})
	{\hat d}^{\dag}_{\mathbf{n}is\sigma}
	{\hat d}^{\phantom{\dag}}_{\mathbf{m}js'\sigma}.
\end{eqnarray}
In this expression
${\hat d}^{\dag}_{\mathbf{n}is\sigma}$
(${\hat d}^{\phantom{\dag}}_{\mathbf{n}is\sigma}$)
are the creation (annihilation) operators of the electron with spin
$\sigma$ at the unit cell
$\mathbf{n}$
in the layer
$i$\,($=1,2$)
in the sublattice
$s$\,($={\cal A,B}$).
For intra-layer hopping, only the nearest-neighbor term is included. Its
value is
$t=-2.57$\,eV.
The inter-layer hoppings are parameterized as described in
Refs.~\onlinecite{TramblyTB_Loc,Trambly2},
with the largest inter-layer hopping amplitude being equal to
$t_0=0.4$\,eV.

Switching to the momentum representation, one can introduce new
single-particle operators
\begin{eqnarray}
{\hat d}^{\phantom{\dag}}_{\mathbf{pG}is\sigma}
=
{\cal N}^{-1/2}\sum_{\mathbf{n}}
e^{-i(\mathbf{p}+\mathbf{G})\mathbf{r}_{\mathbf{n}}^{i}}
{\hat d}_{\mathbf{n}is\sigma}.
\end{eqnarray}
Here
${\cal N}$
is the number of graphene unit cells in the sample in one layer, the
momentum
$\mathbf{p}$
lies in the first Brillouin zone of the superlattice, while
$\mathbf{G}=m_1\bm{{\cal G}}_1+m_2\bm{{\cal G}}_2$
is the reciprocal vector of the superlattice lying in the first Brillouin
zone of the $i$th~layer. The number of such vectors
$\mathbf{G}$
is equal to
$N_{\rm sc}$
for each graphene layer. Thus,
${\hat H}_0$
becomes
\begin{eqnarray}
{\hat H}_0
\!=\!\!\!
\sum_{\mathbf{p}\mathbf{G}_{1,2}}
\sum_{ijss' \sigma }\!
	\tilde{t}_{ij}^{ss'}
	(\mathbf{p}\!+\!\mathbf{G}_1;\mathbf{G}_1 \!-\!\mathbf{G}_2)
	{\hat d}^{\dag}_{\mathbf{pG}_1is\sigma}
	{\hat d}^{\phantom{\dag}}_{\mathbf{pG}_2js'\sigma},\
\label{H0}
\end{eqnarray}
where the hopping amplitudes in momentum space are
\begin{equation}
\label{tG}
\tilde{t}_{ij}^{ss'}(\mathbf{k};\mathbf{G})
\!=\!
\frac{1}{N_{\rm sc}}\!
\mathop{{\sum}'}_{\mathbf{nm}}\!
e^{-i\mathbf{k}(\mathbf{r}_{\mathbf{n}}^{i}-\mathbf{r}_{\mathbf{m}}^{j})}
e^{-i\mathbf{G}\mathbf{r}_{\mathbf{m}}^{j}}\,\,
t(\mathbf{r}_{\mathbf{n}}^{is};
\mathbf{r}_{\mathbf{m}}^{js'})\,.
\end{equation}
The summation symbol with prime
$\sum'_{\bf nm}$
implies that
$\mathbf{m}$
runs over sites inside the zero{\it th} supercell, while
$\mathbf{n}$
runs over all sites in the sample.

\begin{figure}[t]
\centering
\includegraphics[width=0.9\columnwidth]{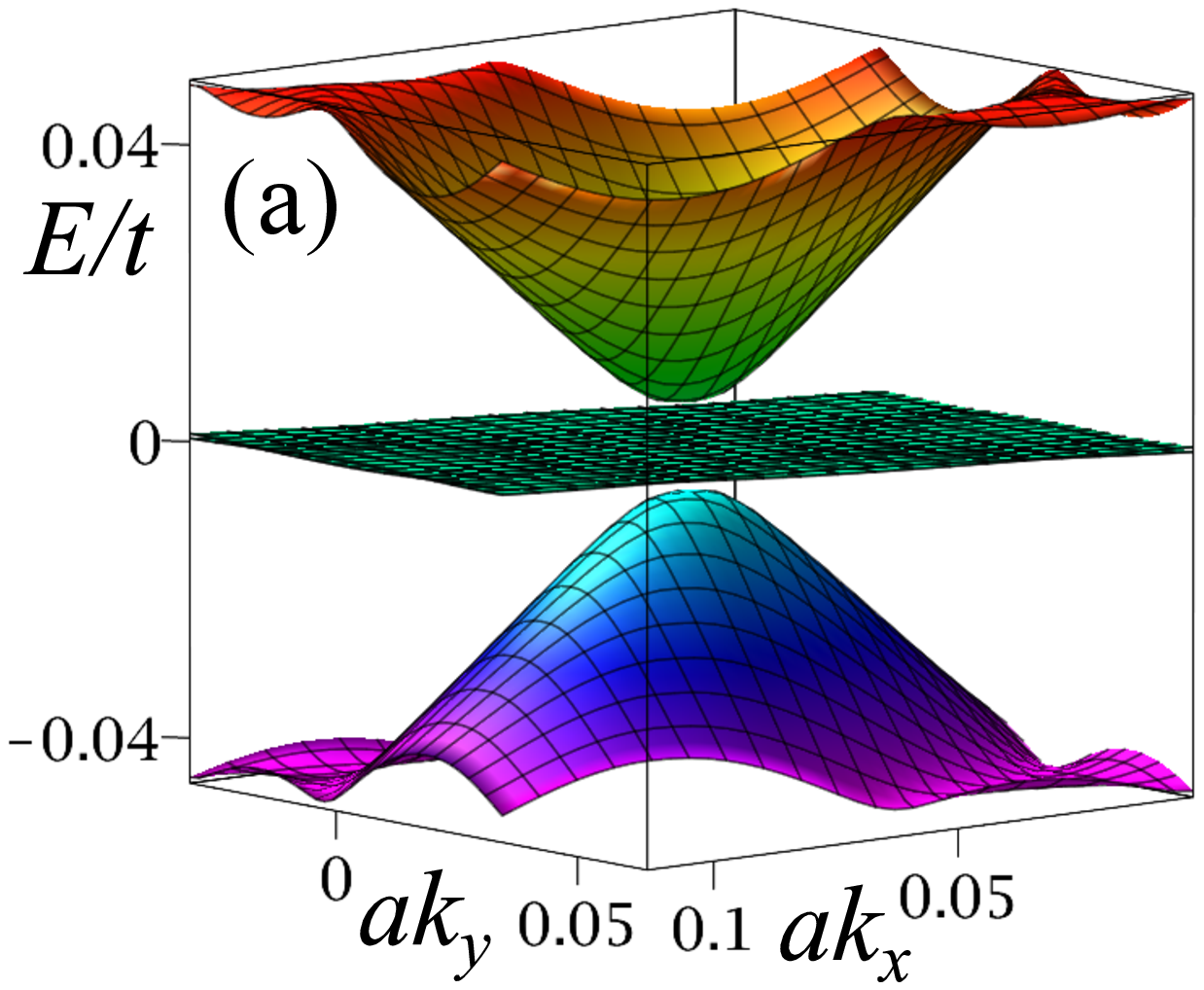}
\includegraphics[width=0.98\columnwidth]{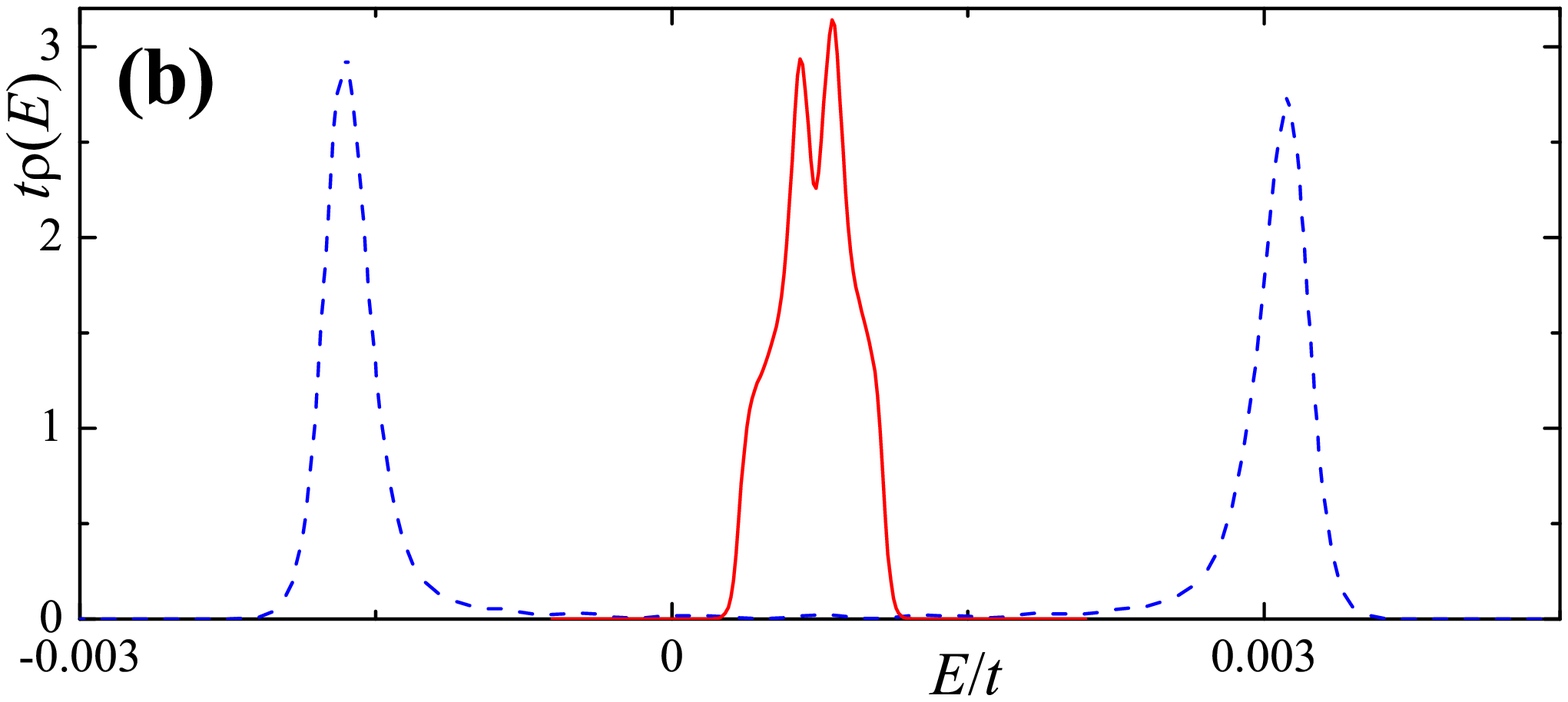}
\caption{(a) Single-particle low-energy band structure (interaction
effects are neglected here) inside the superlattice Brillouin zone
calculated for the first magic angle
$\theta = \theta_c$.
(b)~Low-energy DOS
$\rho(E)$
corresponding to the band structure shown above (solid curve) and for the
band structure modified by interaction (dashed curve), see
Fig.~\ref{fig::spectrumMF}(a)
and text below.
\label{fig::spectrum}
}
\end{figure}
Single-electron energies
$E_{\mathbf{p}}^{S}$
and corresponding eigenvectors
$\Phi^{S}_{\mathbf{pG}i s}$
(here
$S=1,\,2,\,\dots,\,4N_{\rm sc}$
enumerates all
$4N_{\rm sc}$
spin-degenerate bands of the TBLG) are found by numerical diagonalization
of
Eq.~(\ref{H0}).
The spectrum
of~(\ref{H0})
is well-studied. Its properties at small and large $\theta$ differ
qualitatively. When
$\theta> \theta_c$
(for the hopping parameters used here
$\theta_c\approx 1.08^{\circ}$),
the low-energy spectrum is Dirac-like. If
$\theta\leq\theta_c$,
the system acquires a Fermi surface, which is formed by four (eight, if
spin degeneracy is accounted) almost-flat partially degenerate bands at low
energy~\cite{extended_cao_dc}.
In
Fig.~\ref{fig::spectrum}~(a)
the spectrum of this type is plotted for `the first magic angle'
$\theta = \theta_c$.
We see that higher-energy electron and hole bands with pronounced
dispersion are separated from each other by sheets of almost-flat bands.
This peculiar spectrum structure is the origin of the many-body physics
discussed below.
\begin{figure*}[t]
\centering
\includegraphics[width=0.31\textwidth]{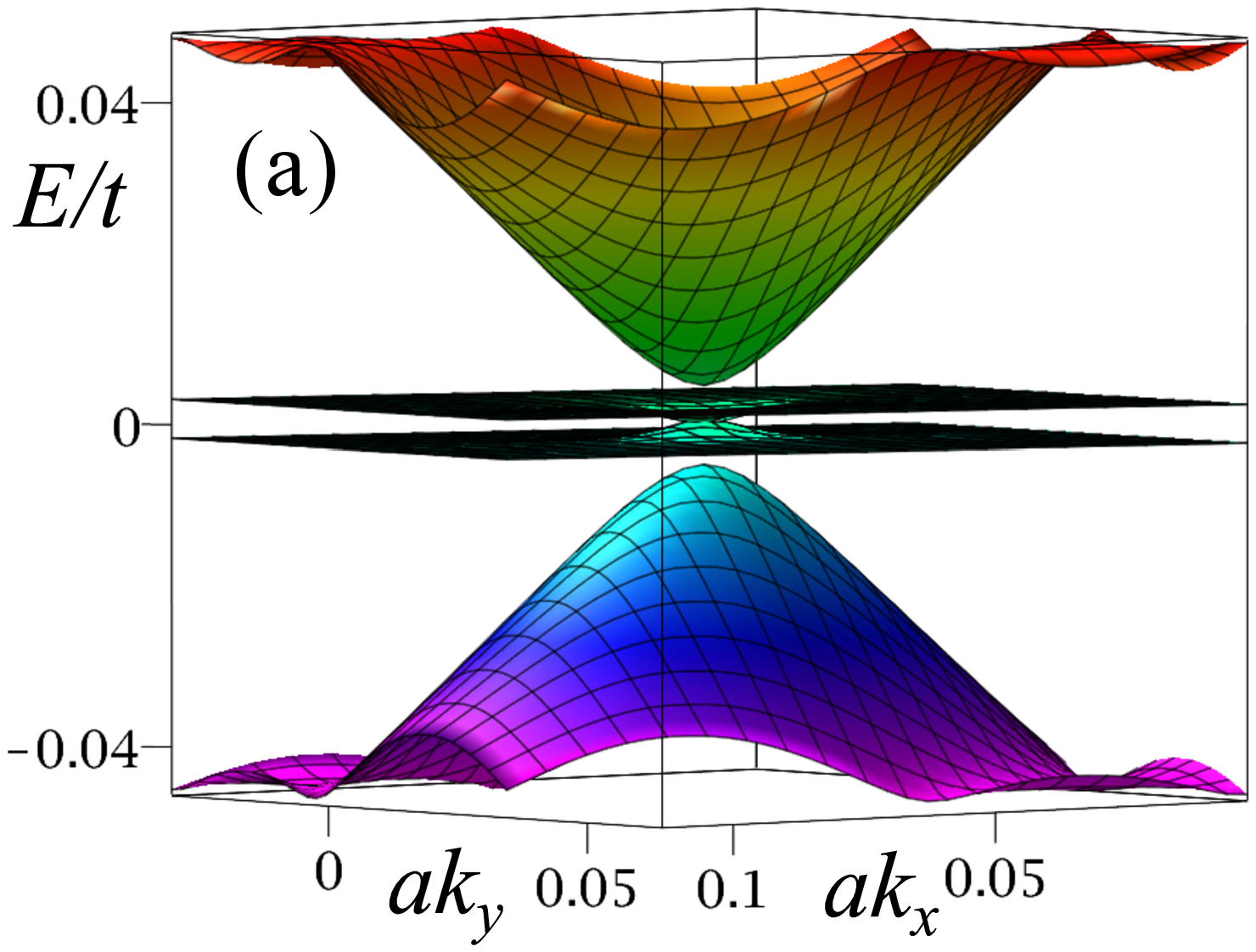}\hspace{1mm}
\includegraphics[width=0.31\textwidth]{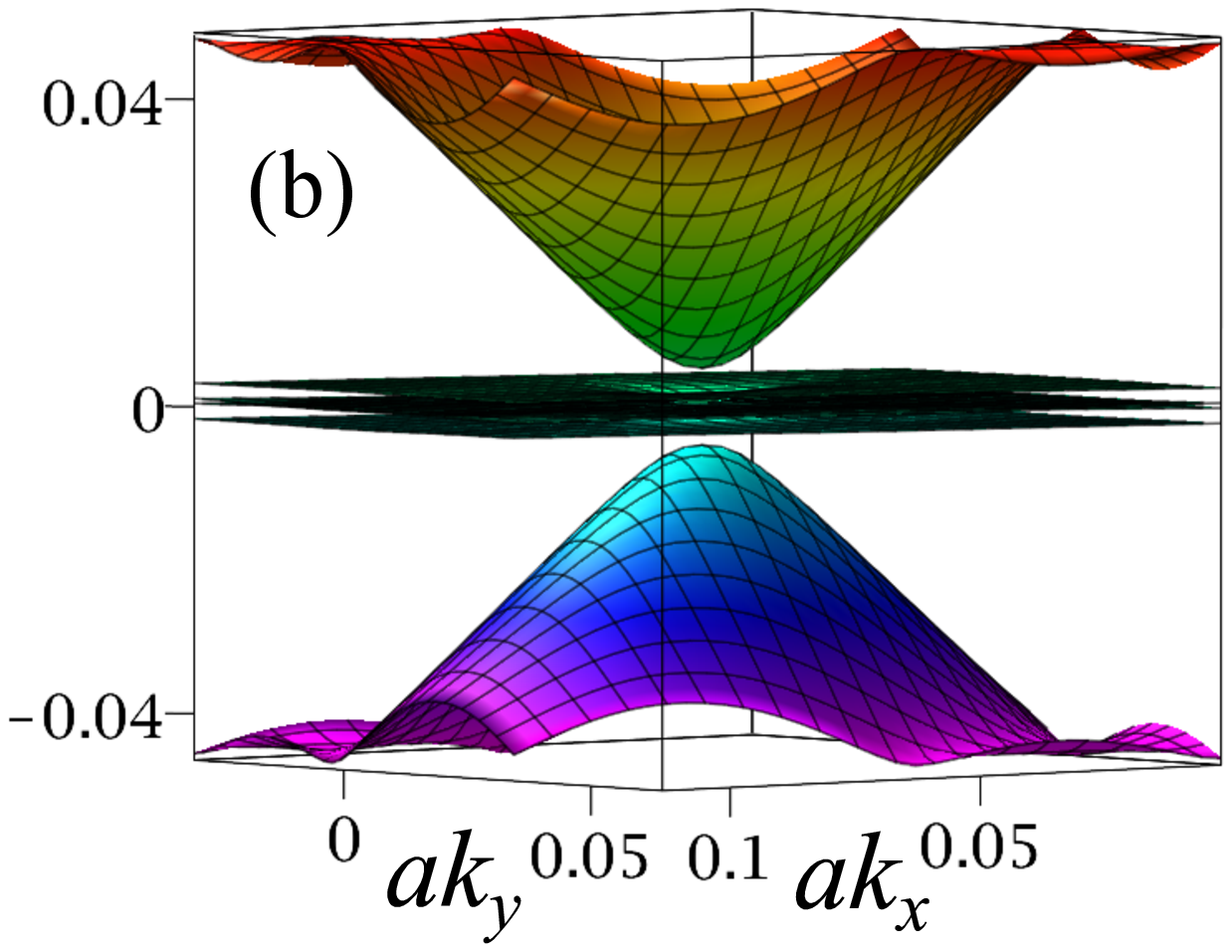}
\includegraphics[width=0.33\textwidth]{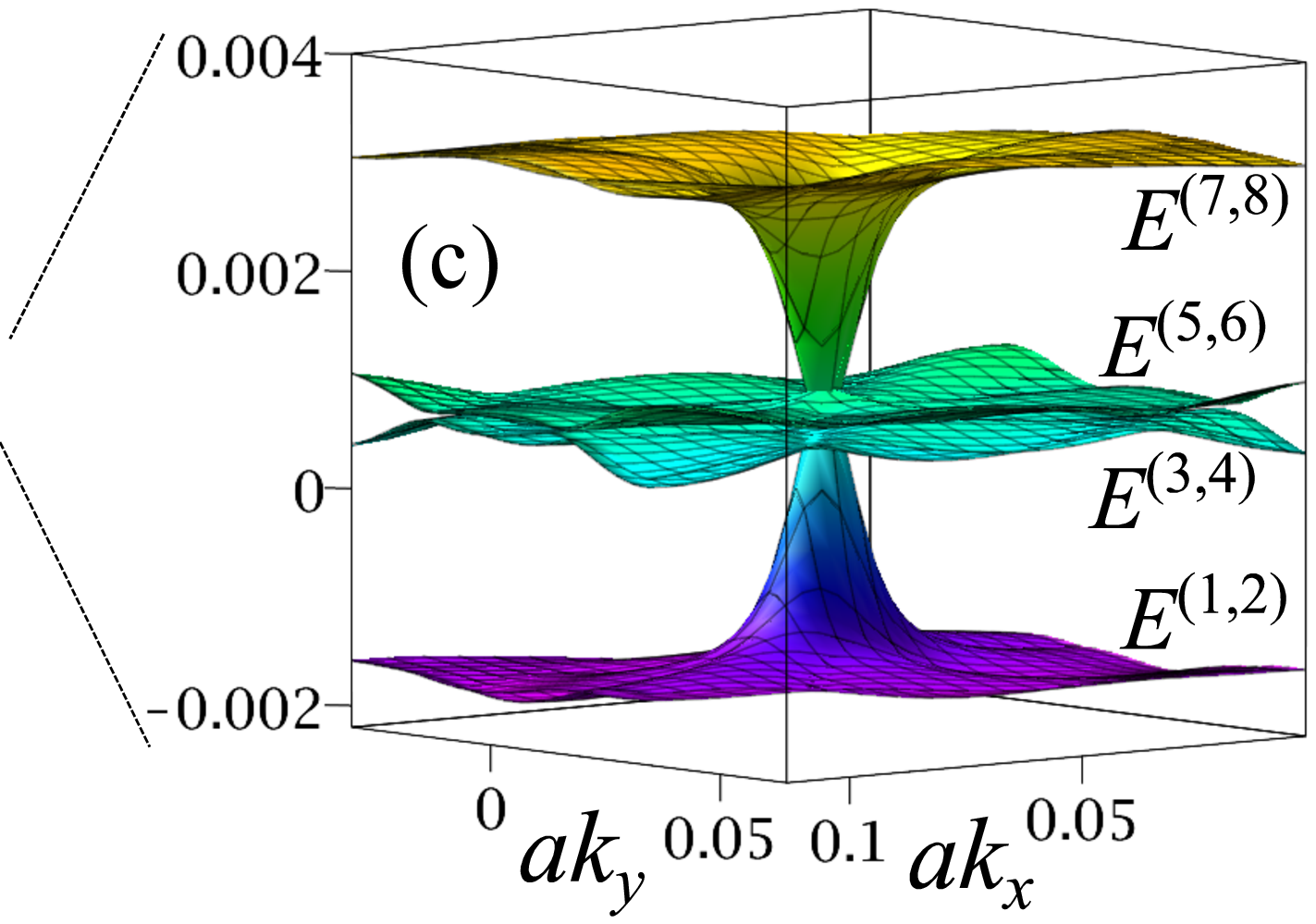}
\caption{(a) Charge neutrality band structure modified by the
interaction~(\ref{U}).
The eight bands are split into two quartets (individual bands are
indiscernible due to small energy separations between the bands of the same
quartet).
(b)~When
$n=n_s/2$,
the order parameters $A$ split the two-quartet structure [panel~(a)] into
the doublet-quartet-doublet structure. (c)~Fine structure of the low-energy
bands shown in panel~(b). The energy bands are labeled by
$E^{(\alpha )}$, $\alpha = 1 \ldots 8$.
\label{fig::spectrumMF}
}
\end{figure*}

To characterize this non-interacting spectrum more thoroughly, it is
instructive to calculate the low-energy DOS
\begin{eqnarray}
\rho (E)
=
2 \sum_S 
	\int \frac{d^2{\bf p}}{ v_{\rm SBZ}} \delta (E^{S}_{\bf p} - E),
\end{eqnarray}
where the integral is taken over the superlattice Brillouin zone, whose
area is denoted by
$v_{\rm SBZ}$.
The DOS is plotted in
Fig.~\ref{fig::spectrum}~(b).
It has a double peak structure, with the total spectral weight
corresponding to eight electrons per a Moir{\'{e}} cell. The DOS remains
non-zero for any doping in the interval
$|n|<n_s$,
as expected for a system with a Fermi
surface~\cite{ourTBLG}.
The Fermi energy for the undoped state
$n=0$
corresponds to the minimum on the DOS plot. The overall structure of the
DOS plot and its width
\begin{eqnarray}
W\sim 2\,{\rm meV}
\end{eqnarray}
are consistent with
Fig.~1d
of
Ref.~\onlinecite{twist_exp_sc2018}.

Numerical calculations demonstrate that the flat bands are separated from
the rest of the spectrum by two gaps, both of the order of 15\,meV, in
qualitative agreement with other computational and
experimental~\cite{twist_exp_sc2018,carr_arXiv2019}
results.

\subsection{Interactions term}

To model experimental
conditions~\cite{twist_exp_insul2018,twist_exp_sc2018},
we study the many-body effects for
$\theta = \theta_c$.
As a starting point of our analysis, we model
${\hat H}_{\rm int}$
using the Hubbard interaction
\begin{eqnarray}
\label{U}
&&{\hat H}_{\rm int}
\!=\!
\frac{U}{2}\!\!
\sum_{{\mathbf{n}is\sigma}} \!\!
	{\hat d}^{\dag}_{\mathbf{n}is\sigma}
	{\hat d}^{\phantom{\dag}}_{\mathbf{n}is\sigma}
	{\hat d}^{\dag}_{\mathbf{n}is\bar{\sigma}}
	{\hat d}^{\phantom{\dag}}_{\mathbf{n}is\bar{\sigma}},\\
\label{eq::Uapprox}
&&
U=2t < U_c^{\rm MF}\,.
\end{eqnarray}
Here the notation
$\bar{\sigma}$
means `not $\sigma$', and
\begin{eqnarray}
\label{eq::Uc_def}
U_c^{\rm MF} \approx 2.23 t
\end{eqnarray}
is the critical strength for a single-layer graphene transition into a
mean-field antiferromagnetic
state~\cite{MF_Uc_sorella1992}.
The
choice~(\ref{eq::Uapprox})
implies that the interaction in our model is strong; yet, not strong enough
to cause a single-layer many-body instability, at least in the mean-field
framework. In other words, the presence of the second layer is a necessary
prerequisite for a mean-field transition.

\section{Mean-field calculations}
\label{sec::MF_calcs}

\subsection{Single-site order parameter}
\label{subsec::Hubbard_OP}

To account for the
interaction~(\ref{U})
at the mean-field level, we must choose a suitable order parameter. First,
let us
define~\cite{guinea_prl2017,PrlOur,PrbROur,PrbOur,AkzyanovAABLG2014}
the single-site magnetization
\begin{eqnarray}
\label{eq::onsite_OP_def}
\eta_{\mathbf{m}is\sigma}
=
\langle {\hat d}^{\dag}_{\mathbf{m}is\sigma}
	{\hat d}^{\phantom{\dag}}_{\mathbf{m}is\bar{\sigma}}
\rangle,
\end{eqnarray}
where
$\langle \ldots \rangle$
denotes the averaging with respect to the mean-field ground state. We will
assume that the anomalous average
$\eta_{\mathbf{m}is\sigma}$,
as a function of position
$\mathbf{m}$,
has the same period as the superlattice. That is, only the spin-rotational
symmetry is broken, while the superlattice translation symmetry is
preserved (the spin texture has the same periodicity as the superlattice).
Using the
$\eta$'s we decouple
$H_{\rm int}$,
to obtain the mean-field interaction
\begin{eqnarray}
\label{eq::OP}
{\hat H}_{\rm int}^{\rm MF}
\!=\!
\sum_{\mathbf{n}is\sigma}\left[
	-\Delta_{\mathbf{n}is\sigma}^{\vphantom{\dagger}}
	{\hat d}^{\dag}_{\mathbf{n}is\bar{\sigma}}
	{\hat d}^{\phantom{\dag}}_{\mathbf{n}is\sigma}
+\frac{\left|\Delta_{\mathbf{n}is\sigma}\right|^2}{2U}\right].
\end{eqnarray}
Here
\begin{eqnarray}
\Delta_{\mathbf{n}is\sigma} = U \eta_{\mathbf{n}is\sigma}
\end{eqnarray}
is the order parameter. Finding the self-consistent value of
$\Delta_{\mathbf{n}is\sigma}$,
we can determine the low-energy band structure of our model, modified by
the
interaction~(\ref{U}).
Figure~\ref{fig::spectrumMF}~(a)
presents the results of such calculations for
$n=0$.
The order parameter
$\Delta_{\mathbf{n}is\sigma}$
lifts the degeneracy of the low-energy spectrum, splitting the eight energy
bands into two quartets: four bands are pushed above the Fermi level
$\varepsilon_{\rm F}$,
and four other bands sink below
$\varepsilon_{\rm F}$.
Each quartet appears as a peak in the DOS plot in
Fig.~\ref{fig::spectrum}(b).
The peaks are separated by
\begin{eqnarray}
\label{eq::charge_neutr_energy_scale}
E_g\approx4.5\times10^{-3}t\approx12\,{\rm meV}.
\end{eqnarray}
Although most of the electronic states are pushed away from the Fermi
energy
$\varepsilon_{\rm F}$,
near the $\Gamma$ point the quartets cross the Fermi energy level, forming
a Fermi surface, and generating a small but finite
$\rho(\varepsilon_{\rm F})$.
Thus, consistent with
experiments~\cite{twist_exp_insul2018},
the undoped state is metallic.

\begin{figure}[t]
\centering
\includegraphics[width=0.99\columnwidth]{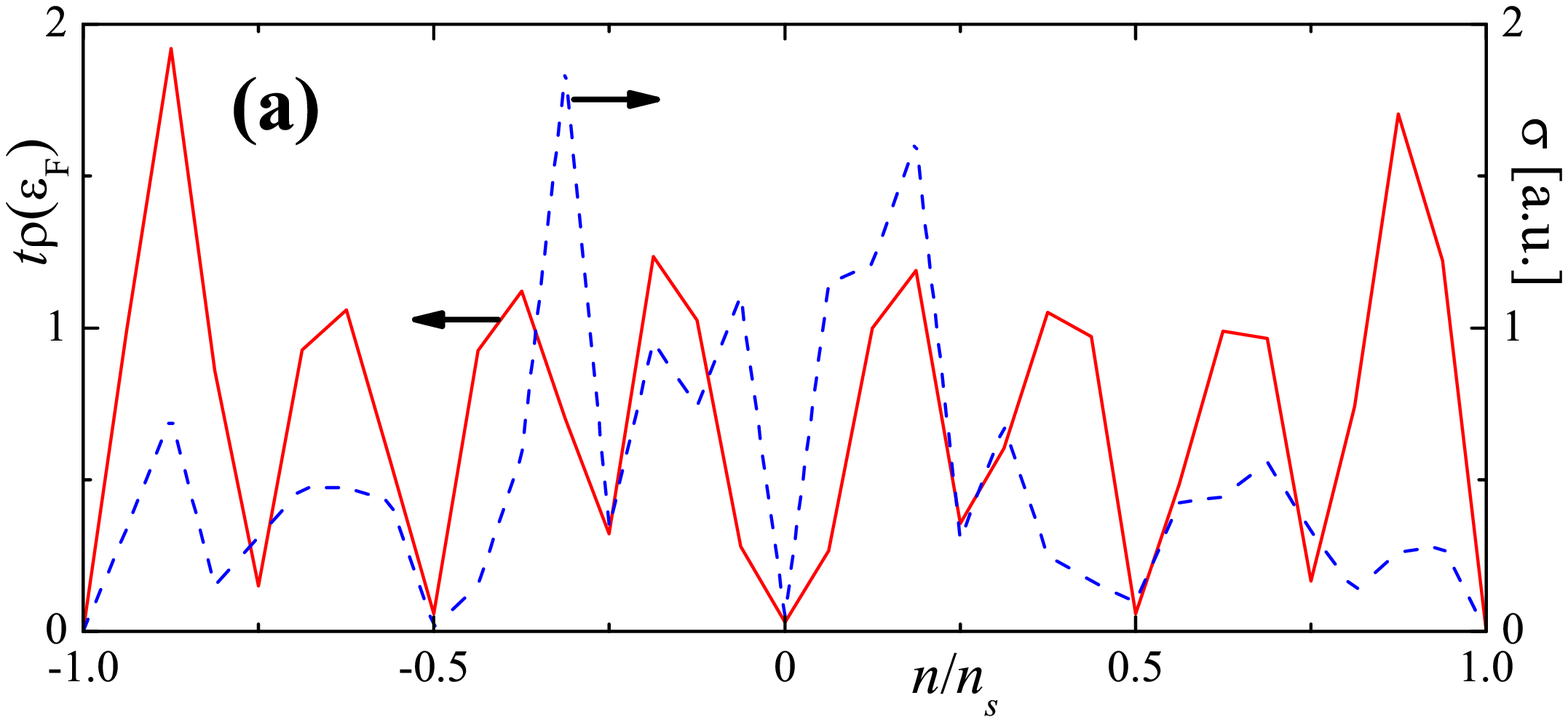}
\includegraphics[width=0.99\columnwidth]{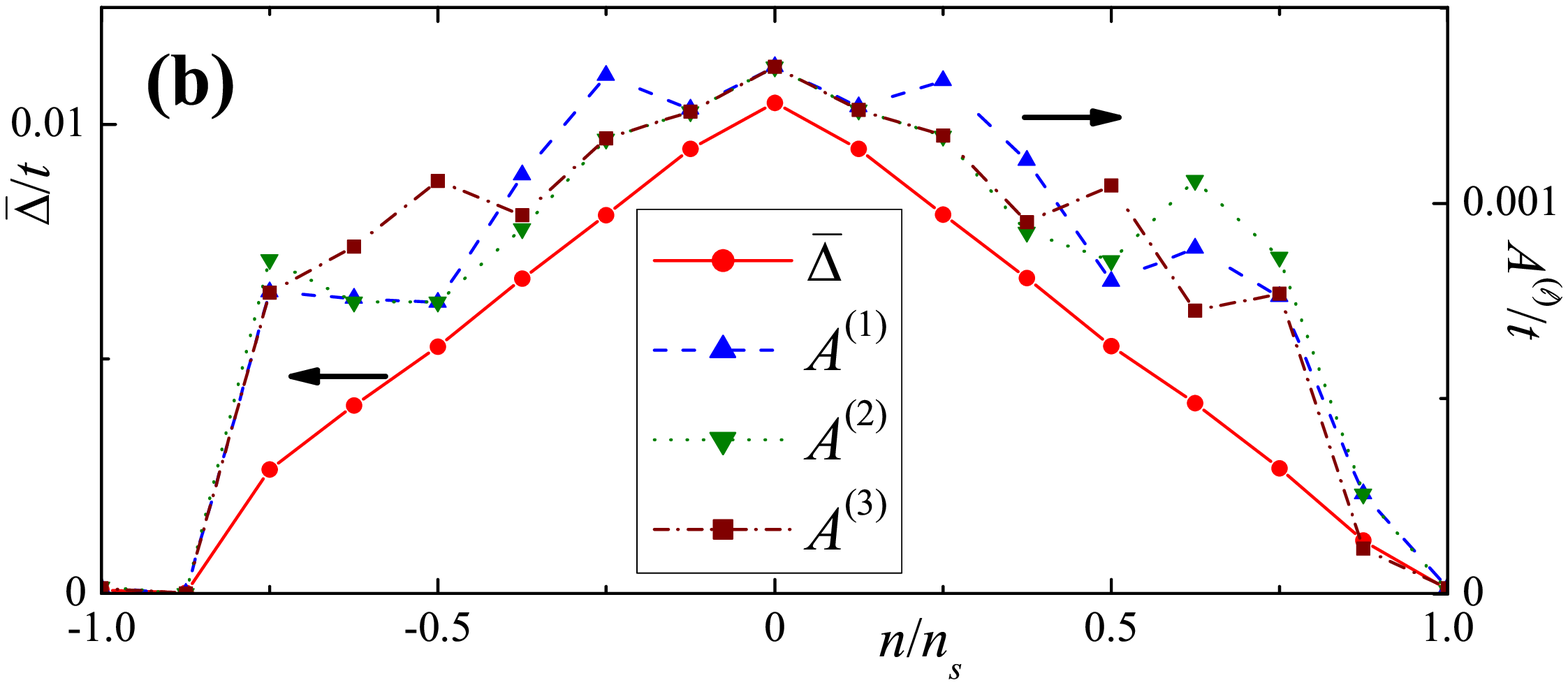}\\
\includegraphics[width=0.99\columnwidth]{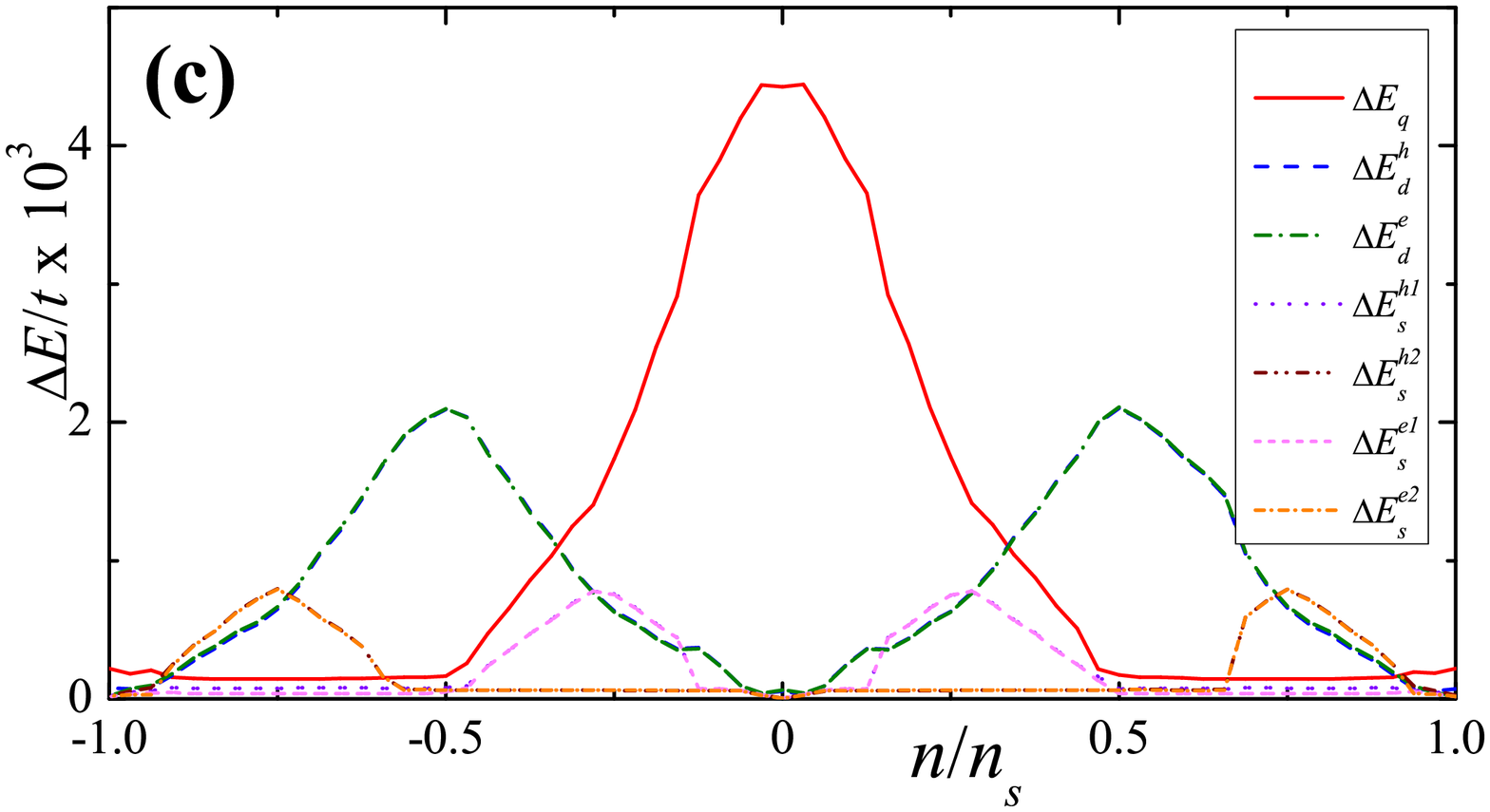}\\
\caption{(a) The DOS at Fermi energy
$\rho ( \varepsilon_{\rm F} )$,
shown by solid curve, and the conductance $\sigma$, shown by
dashed curve, as functions of doping.
(b)~Dependence of
$\bar\Delta$
and
$A^{(\ell)}$,
defined by 
Eqs.~(\ref{eq::Delta_max}-\ref{eq::A3_max}),
on doping $n$. For
$|n| \leq n_s/8$,
we find
$A^{(1)}=A^{(2)} = A^{(3)}$.
When
$|n| > n_s/8$,
the latter identity is violated, indicating the emergence of the so-called
electronic nematicity.
(c)~The dependence of band separation parameters as functions of $n$.
The curves for
$\Delta E_{\rm d}^{\rm e}$
and
$\Delta E_{\rm d}^{\rm h}$
coincide. The same is true for
$\Delta E_{\rm s}^{\rm h1}$
and
$\Delta E_{\rm s}^{\rm e1}$,
as well as for
$\Delta E_{\rm s}^{\rm h2}$
and
$\Delta E_{\rm s}^{\rm e2}$.
\label{fig::DOS_cond_MF}
}
\end{figure}
%

\subsection{Two-site order parameter}
\label{subsec::link_OP}

However, our mean-field calculations show that the order
parameter~(\ref{eq::OP})
is sufficient to describe the conductivity suppression near the charge
neutrality point only. Yet, in the range
$0 < |n| < n_s$
the mean-field theory based on purely single-site order parameter,
Eq.~(\ref{eq::onsite_OP_def}),
predicts quite featureless evolution of the system properties. For our
goals, the most important shortcoming of the purely single-site order
parameter is its inability to split the quartets of the bands further, into
doublets, and single bands.

To appreciate the importance of the latter prerequisite, consider the
following reasoning. Experimentally, doping levels
$n = \pm n_s/2$
are special for the system demonstrates drastic depletion of the
conductivity. On the theory side, doping
$n = n_s/2$
(doping
$n = - n_s/2$)
corresponds to two additional electrons (two additional holes) per
supercell, or, equivalently, it requires complete filling (complete
draining) of exactly two bands of the upper (lower) quartet. Therefore,
an insulating or poorly conducting state at
$n = \pm n_s/2$
requires the separation of the quartet of bands into two doublets, one of
which is filled, the other is empty.

Our numerical study shows that, to generate the desired splitting, the
interaction Hamiltonian, besides the Hubbard
term~\eqref{U},
must include the term describing the (Coulomb) interaction of electrons on
neighboring sites:
\begin{eqnarray}
H^{C}_{\rm int}
=
\frac12\!\!\!
\sum_{{\bf nm}ij\atop ss'\sigma\sigma'}\!\!\!
	V(\mathbf{r}^{is}_{\mathbf{n}}-\mathbf{r}^{js'}_{\mathbf{m}})
	{\hat d}^{\dag}_{\mathbf{n}is\sigma}
	{\hat d}^{\phantom{\dag}}_{\mathbf{n}is\sigma}
	{\hat d}^{\dag}_{\mathbf{m}js'\sigma'}
	{\hat d}^{\phantom{\dag}}_{\mathbf{m}js'\sigma'}.
\quad
\label{eq::H_Coul}
\end{eqnarray}
This interaction can be decoupled by the following excitonic order
parameter:
\begin{eqnarray}
\label{eq::exciton_order}
A^{\mathbf{n}is}_{\mathbf{m}js';\sigma}
=
V(\mathbf{r}^{is}_{\mathbf{n}}-\mathbf{r}^{js'}_{\mathbf{m}})
\langle	
	{\hat d}^{\dag}_{\mathbf{n}is\sigma}
	{\hat d}^{\phantom{\dag}}_{\mathbf{m}js'\bar{\sigma}}
\rangle.
\end{eqnarray}
The mean-field version of the
interaction~\eqref{eq::H_Coul}
is
\begin{eqnarray}
\label{eq::H_CoulMF}
{\hat H}_{\rm int}^{C,\,{\rm MF}}
=
- \frac{1}{2}
\sum_{\mathbf{nm}ij\atop ss'\sigma}\!\!
	\left[
		A^{\mathbf{n}is}_{\mathbf{m}js';\sigma}
		{\hat d}^{\dag}_{\mathbf{m}js'\bar{\sigma}}
		{\hat d}^{\vphantom{\dag}}_{\mathbf{n}is\sigma}
		+ {\rm H.c.}
	\right]
\nonumber\\
+
\frac{1}{2}
\sum_{\mathbf{nm}ij\atop ss'\sigma}\!
\frac{\left|A^{\mathbf{n}is}_{\mathbf{m}js';\sigma}\right|^2}
	{V(\mathbf{r}^{is}_{\mathbf{n}}-\mathbf{r}^{js'}_{\mathbf{m}})}\,.
\end{eqnarray}
For calculations we assume that order parameter
$A^{\mathbf{n}is}_{\mathbf{m}js';\sigma}$
is non-zero only when sites
$\mathbf{r}_{\mathbf{n}}^{is}$
and
$\mathbf{r}_{\mathbf{m}}^{js'}$
are sufficiently close. Namely, if the hopping amplitude connecting
$\mathbf{r}_{\mathbf{n}}^{is}$
and
$\mathbf{r}_{\mathbf{m}}^{js'}$
vanishes, parameter
$A^{\mathbf{n}is}_{\mathbf{m}js';\sigma}$
is zero:
\begin{eqnarray}
\label{eq::finite_order_param}
t(\mathbf{r}_{\mathbf{n}}^{is};\mathbf{r}_{\mathbf{m}}^{js'}) = 0
\Rightarrow
A^{\mathbf{n}is}_{\mathbf{m}js';\sigma} = 0.
\end{eqnarray}
The latter condition implies that for a given site three intra-layer order
parameters
$A^{\mathbf{n}is}_{\mathbf{m}is';\sigma}$,
each associated with a single nearest neighbor, enter the formalism. For a
site on sublattice
${\cal B}$
within a unit cell
${\bf n} = (n,m)$
they are
\begin{eqnarray}
\label{eq::A1}
A_{\mathbf{n}i{\cal B}; \sigma }^{\mathbf{n}i{\cal A}}
&=&
V_{\rm nn}
\langle	
	{\hat d}^{\dag}_{\mathbf{n}i{\cal A}\sigma}
	{\hat d}^{\phantom{\dag}}_{\mathbf{n}i{\cal B} \bar{\sigma}}
\rangle,
\\
\label{eq::A2}
A_{\mathbf{n}i{\cal B};\sigma }^{\mathbf{n}_1i{\cal A}}
&=&
V_{\rm nn}
\langle	
	{\hat d}^{\dag}_{\mathbf{n}_1i{\cal A}\sigma}
	{\hat d}^{\phantom{\dag}}_{\mathbf{n}i{\cal B} \bar{\sigma}}
\rangle,
\\
\label{eq::A3}
A_{\mathbf{n}i{\cal B};\sigma }^{\mathbf{n}_2i{\cal A}}
&=&
V_{\rm nn}
\langle	
	{\hat d}^{\dag}_{\mathbf{n}_2i{\cal A}\sigma}
	{\hat d}^{\phantom{\dag}}_{\mathbf{n}i{\cal B} \bar{\sigma}}
\rangle.
\end{eqnarray}
Here
${\bf n}_1 = (n+1,m)$,
and
${\bf n}_2 = (n,m+1)$.
The nearest-neighbor interaction strength
$V_{\rm nn}$
is equal to
$V_{\rm nn} = V (| \bm{\delta} |)$,
where we take
$V (| \bm{\delta} |)/U=0.59$,
according to
Ref.~\onlinecite{Wehling}.
The quantities defined by
Eqs.~(\ref{eq::A1},\ref{eq::A2},\ref{eq::A3})
satisfy the following relations
\begin{eqnarray}
\left(
	A_{\mathbf{n}i{\cal B}; \sigma }^{\mathbf{n}i{\cal A}}
\right)^*
=
A^{\mathbf{n}i{\cal B}}_{\mathbf{n}i{\cal A}; \bar{\sigma}},
\\
\left(
	A_{\mathbf{n}i{\cal B};\sigma }^{\mathbf{n}_1i{\cal A}}
\right)^*
=
A^{\mathbf{n}i{\cal B}}_{\mathbf{n}_1i{\cal A}; \bar{\sigma}},
\\
\left(
	A_{\mathbf{n}i{\cal B};\sigma }^{\mathbf{n}_2i{\cal A}}
\right)^*
=
A^{\mathbf{n}i{\cal B}}_{\mathbf{n}_2i{\cal A}; \bar{\sigma}},
\end{eqnarray}
which can be verified with the help of
Eq.~(\ref{eq::exciton_order}).

When
$i \ne j$,
quantities
$A^{\mathbf{n}is}_{\mathbf{m}js';\sigma}$
represent inter-layer order parameters. Unlike intra-layer order
parameters,
condition~(\ref{eq::finite_order_param})
does not allow for simple description of non-zero
$A^{\mathbf{n}is}_{\mathbf{m}js';\sigma}$
if
$i \ne j$.
Depending on location of
${\bf r}_{\mathbf{n}}^{is}$
within a supercell,
Eq.~(\ref{eq::finite_order_param})
may allow for as many as $9$
non-vanishing
$A^{\mathbf{n}is}_{\mathbf{m}js';\sigma}$.
Our numerical calculations demonstrate that the inter-layer order
parameters are small, and we will not discuss them in much detail.

The resultant mean-field Hamiltonian equals to
\begin{eqnarray}
H^{\rm MF}
=
H_0 + H_{\rm int}^{\rm MF} + H_{\rm int}^{C,{\rm MF}}.
\end{eqnarray}
It depends on $\Delta$ and $A$. Diagonalizing 
$H^{\rm MF}$,
one finds mean-field eigenenergies
$\tilde{E}^{S}_{\bf p}$,
and total mean-field energy
\begin{eqnarray}
\label{eq::mean_field_energy}
E^{\rm MF} [A, \Delta]
=
\sum_{S {\bf p}}
        \Theta (\varepsilon_{\rm F} -\tilde E^S_{\bf p}) \tilde E^S_{\bf p}
+
\sum_{\mathbf{n}is\sigma}
\frac{\left|\Delta_{\mathbf{n}is\sigma}\right|^2}{2U}
\\
\nonumber 
+
\frac{1}{V_{\rm nn}}
\sum_{\mathbf{n}i\sigma}
\left(
	{|A_{\mathbf{n}i{\cal B}; \sigma }^{\mathbf{n}i{\cal A}}|^2}
	+
	{|A_{\mathbf{n}i{\cal B}; \sigma }^{\mathbf{n}_1i{\cal A}}|^2}
	+
	{|A_{\mathbf{n}i{\cal B}; \sigma }^{\mathbf{n}_2i{\cal A}}|^2}
\right), 
\end{eqnarray}
where the chemical potential
$\varepsilon_{\rm F}$
is chosen such that
\begin{equation}
\frac{4n}{n_s} = \frac{N_{sc}}{{\cal N}} 
\sum_{S{\bf p}}
        \Theta (\varepsilon_{\rm F} - \tilde E^S_{\bf p}) - 4 N_{sc}.
\end{equation}
In principle, both $\Delta$'s and $A$'s can be found executing numerical
minimization of
$E^{\rm MF} [A, \Delta]$
at fixed $n$. Yet, due to large number of sites in a single supercell
($4N_{\rm sc} = 11164$),
straightforward minimization incurs prohibitively high computational costs,
and we have to resort to a simplification. As we will see below, the order
parameter is more than two orders of magnitude smaller than the graphene
band width. Therefore, of all the electronic states of the TBLG, only a
fraction affects significantly the ordering transition: the relevant states
are those whose eigenenergies are close to the Fermi level. All other
states may be accounted perturbatively. To implement this approach, we
project our mean-field Hamiltonian on the subspace spanned by the
eigenvectors
$\Phi^{S}_{\mathbf{pG}i s}$
satisfying the relation:
\begin{eqnarray}
\label{eq::num_window}
-0.25t<\tilde{E}_{\mathbf{p}}^{S}<0.25t\,.
\end{eqnarray}
We then assume that
\begin{eqnarray}
E^{\rm MF} [A, \Delta]
\approx
E^{\rm MF}_{\rm proj} [A, \Delta]
+
\delta E [A, \Delta]
+
{\rm const.},
\end{eqnarray}
where the constant term is independent of $A$ and $\Delta$. The mean-field
energy of the projected Hamiltonian
$E^{\rm MF}_{\rm proj} [A, \Delta]$
is evaluated using the expression identical to
Eq.~(\ref{eq::mean_field_energy})
in which the summation over index $S$ is restricted by
Eq.~(\ref{eq::num_window}).
The contribution from the bands outside
window~(\ref{eq::num_window})
is accounted for by the term
\begin{eqnarray}
\delta E [A, \Delta]
=
-
\frac{\chi_{\rm s}}{2}
\sum_{\mathbf{n}is \sigma }
        \left|\Delta_{\mathbf{n}is \sigma }\right|^2
\\
\nonumber 
-
\chi_{\rm is} \!\!
\sum_{\mathbf{n}i\sigma}
\left(
	{|A_{\mathbf{n}i{\cal B}; \sigma }^{\mathbf{n}i{\cal A}}|^2}
	+
	{|A_{\mathbf{n}i{\cal B}; \sigma }^{\mathbf{n}_1i{\cal A}}|^2}
	+
	{|A_{\mathbf{n}i{\cal B}; \sigma }^{\mathbf{n}_2i{\cal A}}|^2}
\right).
\end{eqnarray}
In this equation
$\chi_{\rm s}$
is the susceptibility of a single-layer graphene to the single-site order
parameter $\Delta$. The susceptibility to the two-site order parameter $A$
is
$\chi_{\rm is}$.
In the limit of the spatially homogeneous antiferromagnetic $\Delta$, it is
known~\cite{MF_Uc_sorella1992}
that
$\chi_{\rm s} = 1/U_c^{\rm MF}$,
see
Eq.~(\ref{eq::Uc_def}).
While
$\chi_{\rm is}$
is not known exactly, we approximate
$\chi_{\rm is} \approx1/U_c^{\rm MF}$.
Since the value of $A$ is very small, the precise value of
$\chi_{\rm is}$
is not crucial.

Applying the described numerical approach, we determined both $A$ and
$\Delta$ for doping in the range
$-n_s<n<n_s$.
To characterize the dependence of the single-site order parameter as a
function of doping, we define
\begin{eqnarray}
\label{eq::Delta_max}
\bar\Delta = \max(|\Delta_{\mathbf{n}is\sigma}|),
\end{eqnarray}
where maximum is taken over a supercell. Similar to
Eq.~(\ref{eq::Delta_max}),
the evolution of the two-site order parameters with doping $n$ can be
characterized by the three quantities defined as follows
\begin{eqnarray}
\label{eq::A1_max}
A^{(1)}
&=&
\max (|A_{\mathbf{n}i{\cal B}; \sigma }^{\mathbf{n}i{\cal A}}|),
\\
\label{eq::A2_max}
A^{(2)}
&=&
\max (|A_{\mathbf{n}i{\cal B};\sigma }^{\mathbf{n}_1i{\cal A}}|),
\\
\label{eq::A3_max}
A^{(3)}
&=&
\max (|A_{\mathbf{n}i{\cal B};\sigma }^{\mathbf{n}_2i{\cal A}}|).
\end{eqnarray}
Each
$A^{(\ell)}$,
$\ell = 1,2,3$,
represents the strength of the order parameter on a specific set of C-C
bonds. Namely,
$A^{(1)}$
describes the order parameters on the bonds which are parallel (or almost
parallel) to
$\bm{\delta}$.
The bonds parallel (or almost parallel) to direction
$(1,\pm\sqrt{3})$
are characterized by
$A^{(2,3)}$.

The plots of
$\bar\Delta$
and
$A^{(\ell)}$
are shown in
Fig.~\ref{fig::DOS_cond_MF}(b).
They demonstrate that the order parameters weaken for larger $n$. Yet,
the doping dependence is not necessary monotonic. In addition, we notice
that, for sufficiently large
$|n|$,
parameters
$A^{(\ell)}$
are no longer equal to each other. In other words, away from the
$n=0$
state the low-energy spectrum spontaneously loses hexagonal symmetry,
indicating the emergence of electronic nematicity. This theoretical
conclusion is consistent with recent experimental
claims~\cite{dos_split2018}.

\subsection{Mean field spectrum structure}
\label{subsec::MF_spectrum}

Once the order parameters are known, we determine the low-energy spectrum
and calculate the DOS at the Fermi level
$\rho (\varepsilon_{\rm F})$
versus $n$, see
Fig.~\ref{fig::DOS_cond_MF}\,(a).
All minima of the DOS occur when $|n|$ is a multiple of
$n_s/4$,
that is, when the doping corresponds to the integer number of electrons per
Moir{\'{e}} cell. The spectrum itself, as function of $n$, experiences
pronounced transformations: depending on $n$, the eight single-particle
bands demonstrate various degeneracy patterns which affect experimentally
measurable quantities, such as
$\rho (\varepsilon_{\rm F})$.

To discuss the specifics of the low-energy spectrum structure, we introduce
index
$\alpha =1\ldots8$,
which, for every momentum
${\bf p}$,
labels the mean-field low-energy eigenstates
$\Phi^{(\alpha)}_{\mathbf{pG}i s}$
according to their mean-field eigenenergies
$E^{(\alpha)}_{\bf p}$
as follows:
$E^{(1)}_{\bf p}<E^{(2)}_{\bf p}<\ldots<E^{(8)}_{\bf p}$.
The detailed structure of this eigenenergy sequence is different for
different $n$. Namely, when
$n=0$,
one has:
\begin{eqnarray}
\label{eq::quartet_ineq}
E^{(1)}_{\bf p}
\approx
E^{(2)}_{\bf p}
\approx
E^{(3)}_{\bf p}
\approx
E^{(4)}_{\bf p}
<
E^{(5)}_{\bf p}
\approx
E^{(6)}_{\bf p}
\approx
E^{(7)}_{\bf p}
\\
\nonumber
\approx
E^{(8)}_{\bf p}.
\end{eqnarray}
In other words, the mean-field spectrum can be described in terms of two
quartets of the single-particle bands: the upper quartet is composed of the
bands
$\alpha = 5, \ldots, 8$,
the bands
$\alpha = 1, \ldots, 4$
belong to the lower quartet, see
Fig.~\ref{fig::spectrumMF}~(a).
The degeneracy within a given quartet is not perfect, yet, the energy
difference between the bands in different quartets is much larger than the
intra-quartet energy separations. The emergence of the quartets is mainly
controlled by the single-site SDW order parameter, as discussed in
subsection~\ref{subsec::Hubbard_OP}.

To quantify the separation between two quartets, we introduce the following
doping-dependent parameter
\begin{eqnarray}
\label{eq::E_AFM}
\Delta E_{\rm q}
=
\int \frac{d^2\mathbf{p}}{v_{\rm SBZ}}
	\left[E^{(5)}_{\mathbf{p}}-E^{(4)}_{\mathbf{p}}\right].
\end{eqnarray}
Non-zero
$\Delta E_{\rm q}$
must not be confused with the gap. Indeed, it is easy to check that, if finite gap
$\delta E = \min_{{\bf p}} \left[ E^{(5)}_{\bf p} - E^{(4)}_{\bf p} \right]$
separating the quartets do exists, then it satisfies
$\delta E < \Delta E_{\rm q}$,
however, finite
$\Delta E_{\rm q}$
coexisting with vanishing
$\delta E = 0$
(as in our case) is also possible.

The dependence of
$\Delta E_{\rm q}$
versus $n$ is plotted in
Fig.~\ref{fig::DOS_cond_MF}\,(c).
We see that the quartet separation is the largest near the charge
neutrality, and virtually zero for
$|n|> n_s/2$.
Near the charge neutrality, the lower quartet is almost entirely filled,
the upper quartet is almost entirely empty. The DOS at the Fermi energy is
finite, but severely depressed, see
Figs.~\ref{fig::spectrum}\,(b)
and~\ref{fig::DOS_cond_MF}\,(a).

The nullification of
$\Delta E_{\rm q}$
for
$|n|> n_s/2$
implies that, when
$n \approx \pm n_s/2$,
the spectrum cannot be described, even approximately, in terms of the upper
and lower quartets. Our numerical calculations demonstrate that for such
doping values each quartet separates into two doublets. The splitting into
the doublets is controlled by the two-site order parameter. 

To characterize the splitting between the doublets, we define
\begin{eqnarray}
\label{eq::E_ex_electr}
\Delta E_{\rm d}^{\rm e}
=
\int \frac{d^2\mathbf{p}}{v_{\rm SBZ}}
	\left[E^{(7)}_{\mathbf{p}}-E^{(6)}_{\mathbf{p}}\right],
\\
\label{eq::E_ex_hole}
\Delta E_{\rm d}^{\rm h}
=
\int \frac{d^2\mathbf{p}}{v_{\rm SBZ}}
	\left[E^{(3)}_{\mathbf{p}}-E^{(2)}_{\mathbf{p}}\right].
\end{eqnarray}
Parameter
$\Delta E_{\rm d}^{\rm e}$
represents the separation of the upper quartet into two doublets, while
$\Delta E_{\rm d}^{\rm h}$
plays the same role for the lower quartet. The splittings
$\Delta E_{\rm d}^{\rm e,h}$
are nearly identical for all $n$'s:
\begin{eqnarray}
\label{eq::spectrum_symm}
\Delta E_{\rm d}^{\rm e} \approx \Delta E_{\rm d}^{\rm h}.
\end{eqnarray}
This feature is sensitive to the specific choice of inter-layer tunneling:
we will demonstrate in another publication that
Eq.~(\ref{eq::spectrum_symm})
is violated, at least at some values of $n$, for different parametrization
of the inter-layer hopping amplitudes.

At
$n = \pm n_s/2$,
the quantities
$\Delta E_{\rm d}^{\rm e,h}$
reach their maximum value:
\begin{eqnarray}
\label{eq::half_fill_separat}
\Delta E_{\rm d}^{\rm e,h} \approx 5\,{\rm meV},
\end{eqnarray} 
while the splitting
$\Delta E_{\rm q}$
becomes small. Therefore, two doublets 
[$E^{(3,4)}$
and
$E^{(5,6)}$]
merge into a quartet, and the whole low-energy bands structure can be
characterized schematically as a doublet-quartet-doublet [see
Fig.~\ref{fig::spectrumMF}~(b,c)].
For
$n = n_s/2$,
the Fermi energy lies between the quartet and the upper doublet. When
$n = - n_s/2$,
the Fermi energy is between the lower doublet and the quartet. Although
there is no well-defined gap between the quartet and either doublets, the
energy separation between the bands is sufficiently strong to induce
pronounced DOS minima at
$n = \pm n_s/2$,
see
Fig.~\ref{fig::DOS_cond_MF}\,(a).

Finally, we want to  discuss the DOS minima at
$n=\pm n_s/4$
and
$n=\pm3n_s/4$.
Since a band quartet accommodates
$n_{\rm s}$
electrons, while a doublet holds
$n_s/2$
electrons, a feature at
$n=\pm n_s/4$,
or at
$n=\pm3n_s/4$
cannot be explained in terms of filling or draining of integer number of
doublets or quartets. As one might guess, such a feature must be associated
with filling or draining of odd number of non-degenerate bands. To enable
the filling or draining of odd number of bands, at least one doublet or
quartet must split into individual bands. To demonstrate the emergence of
non-degenerate singlets in our mean-field formalism, we introduce, similar
to
Eqs.~(\ref{eq::E_AFM}),
(\ref{eq::E_ex_electr}),
and~(\ref{eq::E_ex_hole}),
the following quantities:
\begin{eqnarray}
\Delta E_{\rm s}^{\rm h1,h2}
=
\int\frac{d^2\mathbf{p}}{v_{\rm SBZ}}
	\left[E^{(4,2)}_{\mathbf{p}}-E^{(3,1)}_{\mathbf{p}}\right],
\\
\Delta E_{\rm s}^{\rm e1,e2}
=
\int \frac{d^2\mathbf{p}}{v_{\rm SBZ}}
	\left[E^{(6,8)}_{\mathbf{p}}-E^{(5,7)}_{\mathbf{p}}\right].
\end{eqnarray}
This set of parameters characterizes a separation of a specific band from
the rest of the spectrum.

The dependence of
$\Delta E_{\rm s}^{\rm h1,h2}$
and
$\Delta E_{\rm s}^{\rm e1,e2}$
on doping is shown in
Fig.~\ref{fig::DOS_cond_MF}\,(c).
We see from these plots that
$\Delta E_{\rm s}^{\rm h1,h2}$
and
$\Delta E_{\rm s}^{\rm e1,e2}$
satisfy approximate equalities
\begin{eqnarray}
\label{eq::spectrum_symm2}
\Delta E_{\rm s}^{\rm h1} \approx \Delta E_{\rm s}^{\rm e1},
\quad
\Delta E_{\rm s}^{\rm h2} \approx \Delta E_{\rm s}^{\rm e2}.
\end{eqnarray}
These relations are analogous to
Eq.~(\ref{eq::spectrum_symm}).
As we explained above, the validity of
Eq.~(\ref{eq::spectrum_symm})
depends on particulars of the inter-layer hopping amplitudes
parametrization. The same is true for
Eq.~(\ref{eq::spectrum_symm2})
as well.

The plots in
Fig.~\ref{fig::DOS_cond_MF}\,(c)
reveal that
$\Delta E_{\rm s}^{\rm h1}$
and
$\Delta E_{\rm s}^{\rm e1}$
have maxima at
$n=\pm n_s/4$,
while
$\Delta E_{\rm s}^{\rm h2}$
and
$\Delta E_{\rm s}^{\rm e2}$
have maxima at 
$n=\pm3n_s/4$.
This indicates the emergence of single non-degenerate almost filled and
almost empty electron bands in the TBLG spectrum for these doping values.
However, the details of the low-energy spectrum structure at
$|n| = n_s/4$
and at
$|n| = 3 n_s/4$
are non-identical: for
$|n| = n_s/4$,
parameter
$\Delta E_{\rm q}$
is finite, while at
$|n| = 3n_s/4$,
it is zero. Therefore, the properties of states at
$|n| = n_s/4$
differ from the properties of
$|n| = 3n_s/4$
states.

\section{Discussion}
\label{sec::discussion}

We demonstrated above that the electron-electron interactions modify the
low-energy spectrum of the TBLG, affecting such an important property as
the DOS. In this section, we present an informal review of our results and
discuss their connection to the experiment.

\subsection{Heuristic discussion of the doping-induced spectrum
transformation}

Using numerical optimization of the mean-field energy, we calculated the
low-energy spectrum of the TBLG for various $n$'s. Despite
complexity of the numerical procedure, the resultant doping-induced
evolution of the band structure can be explained qualitatively using simple
heuristic argumentation. Straightforward and intuitive interpretation of
the presented results boosts our confidence in their reliability.

Let us start with the spectrum at the charge neutrality point
$n=0$.
In the absence of interaction, the eigenenergies of the eight bands satisfy
the relation:
\begin{eqnarray}
E^{(1)}_{\bf p}
\approx
E^{(2)}_{\bf p}
\approx
\ldots
\approx
E^{(8)}_{\bf p}.
\end{eqnarray}
Once the interaction is accounted for, the latter relation is replaced by
Eq.~(\ref{eq::quartet_ineq}),
which describe mathematically the splitting of the spectrum into two band
quartets caused by the non-zero
$\Delta_{\mathbf{n}is\sigma}$.
The emergence of two separate quartets minimizes the mean-field energy.
Indeed, the single-electron energies 
$E^{(1)}$,
$E^{(2)}$,
$E^{(3)}$,
and
$E^{(4)}$
of the filled quartet sink, reducing the total energy of the system.
Simultaneous growth of
$E^{(5)}$,
$E^{(6)}$,
$E^{(7)}$,
and
$E^{(8)}$
does not affect the total energy, since this quartet is empty. Upon doping,
the gain in energy due to
$\Delta_{\mathbf{n}is\sigma}$
gradually decreases as the extra charges must go to the states in the upper
quartet. Consequently,
$\Delta_{\mathbf{n}is\sigma}$
decreases when $n$ grows. The same it true for hole doping 
$n<0$.

A similar reasoning suggests that for
$n \approx n_s/2$
energy separation between filled doublet
$E^{(5)}$,
$E^{(6)}$
and empty doublet
$E^{(7)}$,
$E^{(8)}$
becomes favorable. This argument can be trivially extended to
$n \approx - n_s/2$
case. Likewise, at
$n \approx \pm  n_s/4$
and
$n \approx \pm 3 n_s/4$,
the splitting of single non-degenerate bands from the rest of the spectrum
also acts to reduce the total mean-field energy.

\subsection{Comparison with experiment}

\subsubsection{Conductance}
\label{subsec::disc_conduct}

Reference~\onlinecite{twist_exp_insul2018}
presents the experimental measurement of conductance for different doping
values. To establish a connection between our theory and the experiment, we
evaluated the direction-averaged conductance $\sigma$ in the
relaxation-time approximation:
\begin{eqnarray}
\label{eq::kubo_eq}
\sigma
=
\frac{e^2}{4\pi^2}\sum_{S}\int {d^2\mathbf{p}}
\left|\frac{\partial \tilde{E}_{\mathbf{p}}^{S}}{\partial\mathbf{p}}\right|^2
\delta(\varepsilon_{\rm F} - \tilde{E}_{\mathbf{p}}^{S})\tau(\mathbf{p}).
\end{eqnarray}
For calculations, a momentum-independent transport scattering time
$\tau(\mathbf{p})= {\rm const.}$
is assumed. This simplification is very crude, and disregards important
effects (e.g., modifications to $\tau$ due to changes in the DOS, or the
anisotropy). More comprehensive discussion of $\sigma(n)$ will be presented
in a different publication. 

Keeping these reservations in mind, let us examine
Fig.~\ref{fig::DOS_cond_MF}(a),
where
$\sigma(n)$,
estimated with the help of
Eq.~(\ref{eq::kubo_eq}),
is plotted. The conductance demonstrates oscillating dependence on $n$.
Minima of
$\sigma(n)$
coincide mostly with the minima of the DOS. The only exceptions to this
rule are (a)~the emergence of a shallow minimum at
$n \approx -n_s/8$
and (b)~the displacement of minima from
$n=\pm 3 n_s/4$
to
$n \approx \pm 0.8 n_s$.

How do these findings compare against the experiment?
Reference~\onlinecite{twist_exp_insul2018}
presented the conductance measurements in the interval
$|n|<n_s$
for two TBLG samples (sample~D1,
$\theta^{(1)} \approx 1.08^\circ$,
and sample~D4,
$\theta^{(4)} \approx 1.16^\circ$).
The conductances of both samples demonstrated minima at
$n=0$
and
$n = \pm n_s/2$.
Beside these, there were sample-specific minima: for sample D1, there is a
minimum~\cite{weak_minimum}
at
$n=n_s/4$;
for sample D4, there are two
minima~\cite{extended_cao}
at
$n = \pm 3 n_s/4$.
In addition, D4 showed a weaker feature at
$n = n_s/4$.
The available data suggests that (i)~a conductance minimum emerges only
when the value of doping is a multiple of
$n_s/4$,
(ii)~{\it in a given sample}, not every value of $n$ consistent with
condition~(i) necessarily hosts a minimum (the minima at
$n \approx \pm 3n_s/4$
are present for D4, but are absent for D1; when
$n \approx n_s/4$,
only D1 demonstrates the minimum), and (iii)~depending on a sample,
conductance at a given minimum may be metallic (all minima of D4 and the
$n=0$
minimum of D1), or insulating (D1 at
$n=\pm n_s/2$).
Our
Fig.~\ref{fig::DOS_cond_MF}\,(a)
is consistent with (i): disregarding a weak minimum at
$n \approx n_s/8$,
all conductance minima can be associated with multiples of
$n_s/4$.
Our preliminary numerical calculations with a different
model~\cite{Tang}
for the inter-layer hopping show that minima at
$\pm n_s/4$
and
$\pm 3n_s/4$
are susceptible to delicate variations of microscopic details, and may
disappear for some model parameters, in agreement with~(ii). The value of
the conductance at a given minimum demonstrates a similar sensitivity,
which makes our theoretical conclusions compatible with~(iii).

\subsubsection{Nematicity}

Our numerical calculations demonstrate that for sufficiently strong doping
$|n| \gtrsim 0.25 n_s$,
the underlying lattice
$C_6$
symmetry is broken down to 
$C_2$,
see
subsection~\ref{subsec::link_OP}
and
Fig.~\ref{fig::DOS_cond_MF}(b).
This signals the emergence of a metallic phase with spontaneously broken
rotational symmetry. Such a phase is called electron 
nematic~\cite{nematic_review2010}.
Experimental claims of the electronic nematicity observation in a TBLG
sample were recently published in
Ref.~\onlinecite{dos_split2018}.

\subsubsection{Energy scales}

It is known that the mean-field calculations routinely overestimate the
energy scales. For graphene systems with spontaneous symmetry breaking this
circumstance was pointed out in
Ref.~\onlinecite{min_pseudo_fm2008},
see also
Ref.~\onlinecite{coulomb_book2005}. 

When we compare our results against the energy scales extracted from the
data of
Ref.~\onlinecite{twist_exp_insul2018},
we observe that our findings suffer from a similar problem. For example,
let us analyze the effect of the in-plane magnetic field on the many-body
state at
$|n| = n_s/2$.
For the in-plane field, the orbital contribution to the Hamiltonian is
absent, and only Zeeman energy is relevant. Theoretically, it is expected
that the Zeeman contribution weakens the many-body phase: the SDW order
parameters hybridize electronic states with the opposite spin projections,
while the magnetic field polarizes spins effectively removing one of the
projections participating in the ordering. To evaluate the magnetic field
$B$ required to destroy the many-body state at
$n=\pm n_s/2$,
one can write the following
$g\mu_{\rm B} B \sim \Delta E^{\rm e,h}_{\rm d}$
where
$\Delta E^{\rm e,h}_{\rm d}$
is used as a measure for the interaction-induced energy scale in vanishing
field. Employing 
Eq.~(\ref{eq::half_fill_separat})
for
$\Delta E^{\rm e,h}_{\rm d}$,
one obtains
$B \sim 40$\,T.
This estimate is about 5 times higher than the experimental value of 8\,T.

Similar relation between the experimental and theoretical scales can be
established for the charge neutrality point. Figure~3\,c of
Ref.~\onlinecite{twist_exp_insul2018}
plots the dependence 
$\sigma = \sigma(n)$
for different temperatures. The data shows that the minimum of
$\sigma(n)$
at 
$n=0$
disappears above 40\,K. If the latter value is interpreted as the
experimental estimate for the energy scale
$E_g$
(the scale responsible for the single-site ordering at the charge
neutrality point), we see that the experimental result is about three time
smaller than the corresponding theoretical
value~(\ref{eq::charge_neutr_energy_scale}).

\section{Conclusions}
\label{sec::conclusion}

Using a mean-field approximation, we demonstrated that the low-energy flat
bands of TBLG in the low-$\theta$ regime are very sensitive to interactions.
Interactions destroy the partial degeneracy between these bands, inducing
non-trivial many-body states, with magnetism and nematicity. The degeneracy
is lifted in a different manner depending on the doping value. This
microscopic feature is manifested macroscopically as a doping-controlled
sequence of the DOS minima, which can be connected to the conductance
minima recently observed
experimentally~\cite{twist_exp_insul2018}.

\begin{acknowledgments}
This work is partially supported by the JSPS-Russian Foundation for Basic
Research joint Project
No.~19-52-50015.
F.N. is supported in part by the MURI
Center for Dynamic Magneto-Optics via the Air Force Office of Scientific
Research (AFOSR) (FA9550-14-1-0040), Army Research Office (ARO) (Grant
No.~W911NF-18-1-0358), Asian Office of Aerospace Research and Development
(AOARD) (Grant No.~FA2386-18-1-4045), Japan Science and Technology Agency
(JST) (the Q-LEAP, Impact program and CREST Grant No.~JPMJCR1676), Japan
Society for the Promotion of Science (JSPS)
(JSPS-FWO Grant No.~VS.059.18N),
RIKEN-AIST Challenge Research Fund, and the John
Templeton Foundation.
A.V.R. is also grateful to the Skoltech NGP Program (Skoltech-MIT joint
project) for additional support.
\end{acknowledgments}


\end{document}